\documentclass[runningheads]{llncs}

\usepackage[T1]{fontenc}

\usepackage{graphicx}
\usepackage{multirow}

\usepackage{color}
\usepackage{url}

\begin{document}

\title{Reinforcement Learning for Software Vulnerability Analysis: A Systematic Review with Emphasis on C/C++ Source Code and Static Analysis}

\titlerunning{RL for Software Vulnerability Analysis}

\author{Bruno Caro V\'asquez\inst{1}\orcidID{0009-0002-4023-6125} \and
Carola Figueroa Flores\inst{2}\orcidID{0000-0001-6454-7679} \and Gast\'on Marquez Ortega\inst{2}\orcidID{0000-0003-0167-5969}}

\institute{Master’s Student in Computer Science, Universidad del B\'io-B\'io, Chile\\
\email{brunocaro1919@gmail.com}
\and
Department of Computer Science and Information Technology, Universidad del B\'io-B\'io, Chile\\
\email{cfigueroa@ubiobio.cl, gmarquez@ubiobio.cl}}

\maketitle              

\begin{abstract}
Vulnerability detection in C/C++ software remains a major security challenge due to code complexity, manual memory management, and the limitations of traditional static analysis. Reinforcement Learning (RL) has emerged as a promising approach, particularly for fuzzing, test generation, program exploration, and, more recently, vulnerability detection and localization. Following PRISMA 2020, we review RL techniques for software vulnerability analysis, focusing on C/C++ source code and static analysis. We identified 21 primary studies (2015–2026) from major scientific databases and complementary searches. We analyze the addressed tasks, algorithms, state–action–reward formulations, code representations, datasets, and evaluation metrics. Results show that 15 studies focus on fuzzing and guided exploration, only 3 on direct vulnerability detection, and just 1 on statement-level localization. Moreover, statically extracted structural representations such as CFGs and ASTs are rarely used as agent states, and benchmarks lack comparability. We propose a task- and formulation-oriented taxonomy and identify a key research gap: the absence of RL agents that use source-code CFGs as states to detect and localize vulnerable nodes.
\keywords{Reinforcement Learning \and Vulnerability detection \and Vulnerability localization \and Static analysis \and C/C++ \and Control Flow Graph \and Fuzzing}
\end{abstract}

\section{Introduction}
Automatic vulnerability detection in software is one of the most relevant problems in software engineering and cybersecurity. In particular, software development in languages such as C and C++ accounts for a large proportion of documented security vulnerabilities due to characteristics such as manual memory management and the absence of automatic runtime verification mechanisms. Vulnerabilities associated with memory corruption, such as buffer overflow, continue to be widely reported in embedded systems, firmware, operating systems, and critical infrastructure. Several studies identify this type of flaw among the most frequent security issues in applications developed in C and C++, especially in resource-constrained environments \cite{Gomes2025}.

Traditional static analysis tools present widely recognized limitations in the literature: high false-positive rates, limited generalization capability when facing previously unseen vulnerability patterns, and strong dependence on expert-defined rules \cite{Gomes2025}. As a consequence, techniques based on \emph{Machine Learning} have been explored as an alternative to improve automatic detection capabilities, highlighting approaches based on structural code representations, flow analysis, and neural models. In parallel, the use of \emph{Deep Learning} has enabled the optimization of vulnerability-related tasks through fuzzing, improving test input generation, seed selection, and the exploration of complex programs \cite{Miao2022}. However, effective vulnerability detection remains an open problem due to the semantic complexity of code and the diversity of existing vulnerability patterns. In this regard, Li et al. \cite{Li2024} point out that even modern approaches face significant challenges in achieving adequate generalization in real-world scenarios.

Within this context, \emph{Reinforcement Learning} (RL) offers a promising alternative by allowing certain security tasks to be modeled as sequential decision-making problems, where an agent learns a policy from rewards associated with the discovery of relevant behaviors. Unlike supervised \emph{Machine Learning} or \emph{Deep Learning}, which typically map a fixed input to a label in a single step, RL is particularly suited to settings that exhibit (i) sequential decisions, (ii) the need for exploration under uncertainty, (iii) delayed or sparse rewards, (iv) an explicit exploration-exploitation trade-off, and (v) the prioritization of paths or nodes that enables the progressive localization of a target. Several vulnerability-analysis tasks share these traits: fuzzing and program exploration require iterative input or seed selection, whereas fine-grained detection can be cast as navigating a code structure to progressively identify the elements responsible for a vulnerability \cite{Jiang2025,Li2024}.

It is important to note, however, that RL does not remove the dependence on supervision, as the reward signal is frequently derived from labeled datasets, sanitizer outputs, coverage, traces, or fuzzing results, so RL reshapes rather than eliminates the labeling problem. Its application to static source-code analysis thus poses challenges in defining the environment, state, reward, and suitable benchmarks.

Building on this motivation, this review synthesizes the state of the art on RL applied to software vulnerability detection, analyzing its tasks, models, and available resources, to assess the feasibility of designing an RL agent oriented toward this task.

To this end, this review makes the following contributions: (i) A systematic mapping of RL-based approaches for software vulnerability analysis, conducted under the PRISMA 2020 guidelines. (ii) A taxonomy organized by task: fuzzing, test generation, program exploration, vulnerability detection, and fine-grained localization. (iii) A comparative analysis of the RL formulations reported in the corpus, in terms of state, action, reward, and environment. (iv) The identification of open gaps for static source-code vulnerability localization based on structural representations such as CFGs or ASTs.

\section{Methodology}
This review follows the PRISMA 2020 statement guidelines \cite{Pagen71}, adapting its structure to the needs of software engineering, cybersecurity, and computer science. Likewise, the planning, search, selection, and evidence synthesis process was based on the guidelines for conducting Systematic Literature Reviews in Software Engineering proposed by Kitchenham and Charters \cite{Kitchenham2007}. The study was designed to investigate proposals, implementations, or evaluations of models based on Reinforcement Learning applied to source-code vulnerability detection, with special emphasis on structural code representations under a static analysis approach.

The process was divided into the four main methodological phases of PRISMA: Identification, Screening, Eligibility, and Inclusion. This ensures transparency, reproducibility, and methodological rigor in the retrieval of potentially relevant articles.

To delimit the research questions and criteria, the PICOC framework was used. In this context, \emph{Population} considers studies on software vulnerabilities, source-code analysis, C/C++ programs, or multi-programming-language vulnerability datasets. \emph{Intervention} focuses on Reinforcement Learning applied to vulnerability analysis, detection, and exploration. \emph{Comparison} considers contrasts with traditional static analysis, supervised and deep learning, conventional fuzzing, and SAST tools. \emph{Outcomes} includes performance and detection metrics such as precision, recall, and generalization. Finally, \emph{Context} focuses on static or hybrid analysis.

\subsection{Research Questions}
This review is guided by the following research questions (RQs), aimed at identifying trends and gaps within the investigated context:

\begin{itemize}
    \item RQ1. What software security tasks have been addressed using Reinforcement Learning in the context of source-code vulnerabilities?

    \item RQ2. Which RL or Deep RL algorithms have been used most frequently in this domain?

    \item RQ3. How are the components of the RL problem defined, specifically states, actions, rewards, and environment?

    \item RQ4. What source-code representations are used to feed or guide RL agents?

    \item RQ5. What datasets, benchmarks, and metrics are used to evaluate these approaches?

    \item RQ6. What limitations, methodological risks, and gaps persist when applying RL to vulnerability detection in C/C++?

\end{itemize}

\subsection{Search Strategy}
To ensure comprehensive coverage of the current related literature, three high-impact databases were used as primary sources: Scopus, Web of Science, and Taylor \& Francis. IEEE Xplore was later added as a complementary database to strengthen coverage of recent venues, and it also contributed primary studies to the corpus. Google Scholar was used only to verify coverage and availability, and backward snowballing (citation searching) was applied as an additional retrieval method, and neither was treated as a primary database. These sources were selected for their availability and their coverage of publications in AI, computer science, and cybersecurity.

Publications from the period between 2015 and 2026 were considered, restricted to peer-reviewed primary studies written in English or Spanish. As the review was conducted during 2026, the search was closed in July 2026, and no records published after that date were included.

For this purpose, a search was conducted using three dedicated search strings. The first string was considered the most precise, as it aimed to better capture the core topic under investigation, whereas the remaining strings provided broader thematic and structural coverage, enabling the identification of other uses of RL related to cybersecurity or structural representation methods used by such models.

\begin{itemize}
    \item \footnotesize
    \parbox[t]{\linewidth}{\raggedright
    \texttt{( "reinforcement learning" OR "deep reinforcement learning" OR "RL agent" OR “DQN” OR “PPO” OR "actor-critic" OR "Q-learning" ) AND ( "software vulnerability" OR "vulnerability detection" OR "code vulnerability" OR "security bug" OR "software security" ) AND ( "source code" OR "static analysis" OR "program analysis" OR "code analysis" OR “AST” OR “CFG” OR “PDG” )}}
    \item \footnotesize
    \parbox[t]{\linewidth}{\raggedright
    \texttt{( "reinforcement learning" OR "deep reinforcement learning" ) AND ( “fuzzing” OR "test generation" OR "program testing" OR "vulnerability discovery" ) AND ( "software security" OR "binary analysis" OR "source code" OR "C/C++" )}}
    \item \footnotesize
    \parbox[t]{\linewidth}{\raggedright
    \texttt{( "reinforcement learning" OR "deep reinforcement learning" ) AND ( "abstract syntax tree" OR AST OR "control flow graph" OR CFG OR "program dependence graph" OR PDG OR "code representation" ) AND ( "vulnerability" OR "software security" OR "bug detection" )}}
\end{itemize}

\subsection{Eligibility Criteria}
Inclusion and exclusion criteria (Table \ref{tab:criterios}) were applied during the screening and eligibility phases.

\begin{table}[ht]
\centering
\caption{Inclusion and exclusion criteria}
\label{tab:criterios}
\resizebox{0.75\linewidth}{!}{%
\begin{tabular}{|p{0.46\linewidth}|p{0.46\linewidth}|}
\hline
\textbf{Inclusion criteria} & \textbf{Exclusion criteria} \\
\hline
Articles published between 2015 and 2026 (+ articles from the last 12 months). & General pentesting studies without analysis of programs, code, binaries, or software vulnerabilities. \\
\hline
Articles written in English or Spanish. & Network intrusion detection studies (IDS/IPS) unrelated to source code or software. \\
\hline
Peer-reviewed primary studies. & Articles without empirical evaluation or with insufficient methodological description. \\
\hline
Studies using RL or Deep RL in tasks related to vulnerabilities, program analysis, fuzzing, software security, or fault detection/prioritization. & Surveys, secondary reviews, books, theses, or grey literature, unless used only as background. \\
\hline
Studies with sufficient description of the algorithm, environment, reward, or evaluation. & Studies where RL is mentioned only superficially and is not a central part of the method. \\
\hline
Studies applied to C/C++ or multi-programming-language datasets, provided there is a clear relationship with software vulnerabilities. & Works focused only on automatic repair if they do not address vulnerability detection, prioritization, exploration, or analysis. \\
\hline
\end{tabular}%
}
\end{table}

Since the inclusion criteria deliberately admit a broad range of RL-based security tasks, the corpus is stratified by its relationship to the review objective. \emph{Core} studies apply RL directly to vulnerability detection or localization over source code. \emph{Peripheral but relevant} studies address adjacent tasks such as fuzzing or concolic execution, which inform the RL formulation but not static source-code detection. \emph{Excluded} works cover network intrusion detection, general pentesting, or settings where RL is not central. This makes explicit that fuzzing- and binary-oriented studies enter as related RL vulnerability-analysis tasks, not as static source-code detection, preserving the alignment with the review's emphasis on C/C++ source code.

\subsection{Study Selection Process}

\begin{figure}[ht]
\centering
\includegraphics[width=0.8\linewidth]{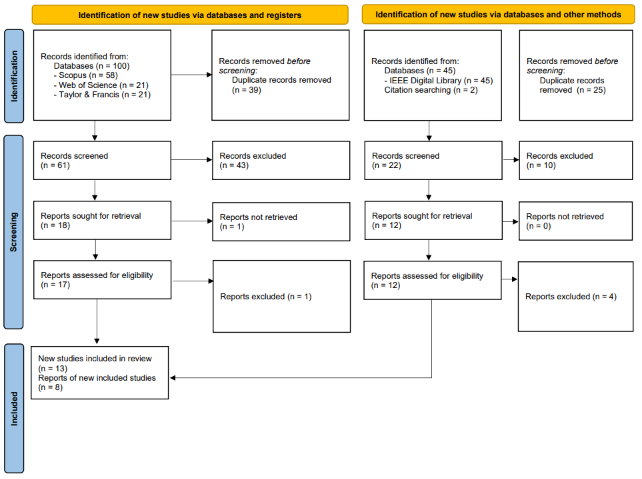}
\caption{PRISMA flow diagram of the study selection process}
\label{fig:prisma}
\end{figure}

During the identification phase, the three search strings were executed across the primary databases, yielding 58 records from Scopus, 21 from Web of Science, and 21 from Taylor \& Francis, amounting to 100 records. The complementary IEEE Xplore search retrieved 45 additional records, and citation searching contributed 2. The per-source flow is summarized in Table \ref{tab:fuentes}, and the complete PRISMA flow in Fig. \ref{fig:prisma}.

During screening, titles and abstracts were assessed against the eligibility criteria: duplicates were removed, as were studies that did not use RL for code analysis or cybersecurity tasks, or lacked a clear description of the RL model/agent. This left 18 articles (13 from Scopus, 5 from Web of Science).

During eligibility, the environment, reward, policy, actions, and metrics of each candidate were examined against the criteria in Table \ref{tab:criterios}. Studies were excluded when the full text was inaccessible through institutional or open-access sources, or when they did not comply with the criteria. The database arm, the complementary IEEE Xplore search, and citation searching produced an initial corpus of 13 primary studies. A later retrieval stage recovered 12 previously inaccessible Scopus and Web of Science records, of which 8 qualified, yielding the final corpus of 21 studies. Per-source counts appear in Table \ref{tab:fuentes} and the overall flow in Fig. \ref{fig:prisma}.

\begin{table}[ht]
\centering
\footnotesize
\caption{Records retrieved and retained per source}
\label{tab:fuentes}
\resizebox{0.9\linewidth}{!}{%
\begin{tabular}{|l|c|c|c|c|c|}
\hline
\textbf{Source} & \textbf{Retrieved} & \textbf{After dup.} & \textbf{Post-Screen} & \textbf{Full-text} & \textbf{Included} \\
\hline
Scopus              & 58 & 34 & 13 & 12 & 09 \\
Web of Science      & 21 & 10 & 05 & 05 & 04 \\
Taylor \& Francis   & 21 & 17 & 00 & 00 & 00 \\
IEEE Xplore (compl.)& 45 & 20 & 10 & 10 & 07 \\
Citation searching  & 02 & 02 & 02 & 02 & 01 \\
\hline
\textbf{Total}      & 147 & 83 & 30 & 29 & \textbf{21} \\
\hline
\end{tabular}%
}
\end{table}

\section{Results}
\subsection{Descriptive Analysis}

A total of 21 primary studies were included. Their publication years span 2019--2026, peaking at 7 in 2024 and 3 in 2025. This reflects the consolidation of deep RL frameworks and growing interest in modeling security tasks as sequential decision-making, rather than a maturation of source-code vulnerability detection, as shown below.

By task, the corpus is dominated by guided fuzzing, addressed by 15 of the 21 studies \cite{Chen2021,Gong2022,Gotz2025,Huang2026,Huang2022,Jhang2024,Khan2025,Kuznetsov2019,Kuznetsov20191,Liang2022,Paduraru2021,Pham2024,Wang2021,Xie2024,Yu2024}. The remainder covers hierarchical seed scheduling\cite{Wang2021}, concolic execution\cite{Paduraru2020}, unit-test generation\cite{Steenhoek2025}, cyber-physical evaluation\cite{Ding2023}, and three direct-detection studies\cite{Jiang2025,Li2024,article}. This imbalance is expected: fuzzing offers sequential actions and coverage-based rewards, whereas static detection depends more on external labels and benchmark design. Table \ref{tab:estudios-seleccionados} distinguishes the \textit{Core} studies (S06, S08, S21) from the remaining \textit{Peripheral} works.

Four algorithmic families appear: value-based RL, policy-gradient RL, multi-armed bandits, and a policy-gradient variant within a deep equilibrium network. DQN and its variants dominate, in eight studies \cite{Chen2021,Gotz2025,Huang2026,Kuznetsov2019,Kuznetsov20191,Liang2022,Paduraru2020,Paduraru2021}, followed by UCB/MAB in about six \cite{Huang2022,Jhang2024,Pham2024,Wang2021,Xie2024,Yu2024}, PPO in three \cite{Khan2025,Paduraru2021,Steenhoek2025}, and REINFORCE/VPG in the detection-oriented works \cite{Jiang2025,Li2024,article}. DQN's dominance reflects the fuzzing bias, as mutation, seed, and operator selection are discrete actions suited to value-based methods, and says little about detection, where objectives and action spaces differ. Conversely, policy-gradient methods concentrate on detection and localization, which optimize decisions over sequences or label predictions rather than discrete operators, favoring stable policy optimization.

\begin{table}[ht]
\centering
\caption{Characterization of the selected studies}
\label{tab:estudios-seleccionados}
\resizebox{0.8\linewidth}{!}{%
\begin{tabular}{c l l l l}
\hline \hline
\textbf{ID} & \textbf{Reference} & \textbf{Task} & \textbf{Algorithm} & \textbf{Scope} \\
\hline \hline
S01 & Paduraru et al.\cite{Paduraru2021} & Binary fuzzing & DQN / PPO & Peripheral \\
\hline
S02 & Kuznetsov et al.\cite{Kuznetsov2019} & Black-box fuzzing & DQN & Peripheral \\
\hline
S03 & Kuznetsov et al.\cite{Kuznetsov20191} & Black-box fuzzing & DQN & Peripheral \\
\hline
S04 & Steenhoek et al.\cite{Steenhoek2025} & Test generation & PPO (RLHF) & Peripheral \\
\hline
S05 & Götz et al.\cite{Gotz2025} & Hardware fuzzing & DQN variants & Peripheral \\
\hline
S06 & Li et al.\cite{Li2024} & Direct detection & REINFORCE/DEQ & Core \\
\hline
S07 & Xie et al.\cite{Xie2024} & Directed fuzzing & UCB bandit & Peripheral \\
\hline
S08 & Ren et al.\cite{article} & Direct detection & VPG & Core \\
\hline
S09 & Pham et al.\cite{Pham2024} & Grey-box fuzzing & UCB bandit & Peripheral \\
\hline
S10 & Chen\cite{Chen2021} & Binary fuzzing & DQN + Intel PT & Peripheral \\
\hline
S11 & Ding et al.\cite{Ding2023} & RAV vuln. evaluation & RL & Peripheral \\
\hline
S12 & Paduraru et al.\cite{Paduraru2020} & Concolic execution & DQN + LSTM & Peripheral \\
\hline
S13 & Wang et al.\cite{Wang2021} & Hierarchical scheduling & UCB/MAB & Peripheral \\
\hline
S14 & Liang and Xiao\cite{Liang2022} & Directed fuzzing & DQN & Peripheral \\
\hline
S15 & Huang et al.\cite{Huang2022} & Kernel fuzzing & MAB & Peripheral \\
\hline
S16 & Jhang and Huang\cite{Jhang2024} & Multi-argument fuzzing & MAB & Peripheral \\
\hline
S17 & Yu et al.\cite{Yu2024} & Directed greybox fuzzing & MAB & Peripheral \\
\hline
S18 & Khan et al.\cite{Khan2025} & Embedded fuzzing & PPO & Peripheral \\
\hline
S19 & Gong et al.\cite{Gong2022} & Format-constrained fuzzing & DDPG & Peripheral \\
\hline
S20 & Huang et al.\cite{Huang2026} & Modbus fuzzing & LABM-DQN & Peripheral \\
\hline
S21 & Jiang et al.\cite{Jiang2025} & Direct detection & REINFORCE & Core \\
\hline \hline
\end{tabular}%
}
\end{table}

Code representation is the most heterogeneous component of the corpus (Table \ref{tab:representaciones-codigo}), spanning from black-box outputs and execution traces to structural representations, the latter almost always derived from binaries. Only S08\cite{article}, S15\cite{Huang2022}, and S21\cite{Jiang2025} operate on source code, and of these only S08 and S21 target detection or localization, underscoring the limited use of static source-code structure in RL-based vulnerability analysis.

\begin{table}[ht]
\centering
\caption{Code representations used}
\label{tab:representaciones-codigo}
\resizebox{0.8\linewidth}{!}{%
\begin{tabular}{l c l c c}
\hline \hline
\textbf{Representation} & \textbf{n} & \textbf{Studies} & \textbf{Code access} & \textbf{Static} \\
\hline \hline
Black box (program output only) & 2 & S02, S03 & No-code-access & No \\
\hline
Binary execution traces (basic blocks) & 2 & S01, S10 & Binary-only & No \\
\hline
Hardware RTL simulation output & 1 & S05 & Firmware/hardware & No \\
\hline
Firmware testing in RAVs & 1 & S11 & Firmware/hardware & No \\
\hline
CFG from binary (IDA Pro + assembler) & 1 & S06 & Binary-only & Yes \\
\hline
Instrumented source basic blocks (ASan + taint) & 2 & S07, S09 & Partial-source & No \\
\hline
Source-code token sequences (LLM prompt) & 1 & S04 & Partial-source & Yes \\
\hline
C/C++ source code at function level & 1 & S08 & Yes-source & Yes \\
\hline
Path constraints (LSTM) & 1 & S12 & Binary-only & No \\
\hline
Multilevel coverage clusters (MAB tree) & 1 & S13 & Binary-only & No \\
\hline
Execution path/distance (binary instrumentation) & 1 & S14 & Binary-only & No \\
\hline
Execution path/distance (embedded firmware) & 1 & S18 & Firmware/hardware & No \\
\hline
CFG from source code (LLVM+SVF, block weights) & 1 & S15 & Yes-source & Yes \\
\hline
Instrumented binary basic blocks (ASan) & 1 & S16 & Binary-only & No \\
\hline
Distance-based CFG/CG + static reports (greybox) & 1 & S17 & Partial-source & Partial \\
\hline
Input data boundaries/format & 1 & S19 & Binary-only & No \\
\hline
Binary bit-flip without source code (protocol) & 1 & S20 & No-code-access & No \\
\hline
Source-code statement embeddings (CodeBERT-HLS) & 1 & S21 & Yes-source & Yes \\
\hline \hline
\end{tabular}%
}
\end{table}

Datasets and benchmarks also present considerable diversity across the corpus. LAVA-M\cite{DolanGavitt2016LAVA} is the most recurrent in fuzzing studies, employed by Xie et al. and Wang et al., together with the DARPA CGC dataset\cite{DARPA_CGC}\cite{Wang2021,Xie2024}. DeepEXE\cite{Li2024} evaluates on Juliet\cite{Juliet13}, NDSS18, FFmpeg, and Esh, whereas source-level detection relies on Reveal\cite{Chakraborty2022Reveal} and FFmpeg+QEMU (S08) and on BigVul\cite{Fan2020BigVul} (S21). Many remaining studies use program-specific, proprietary, hardware-associated, or domain-specific environments.

\begin{table}[t]
\centering
\caption{Datasets and benchmarks used in the selected studies}
\label{tab:datasets-benchmarks}
\resizebox{0.9\columnwidth}{!}{%
\begin{tabular}{p{0.42\columnwidth} p{0.28\columnwidth} p{0.55\columnwidth}}
\hline \hline
\textbf{Dataset / Benchmark} & \textbf{Studies} & \textbf{Type} \\
\hline \hline
LAVA-M\cite{DolanGavitt2016LAVA} & S07, S13, S18 & Synthetic C/C++ injected bugs (fuzzing) \\
\hline
DARPA CGC\cite{DARPA_CGC} & S07, S13 & Semi-synthetic binary (fuzzing) \\
\hline
Google FuzzBench & S13 & Real-world C/C++ fuzzing projects \\
\hline
NDSS18, Juliet\cite{Juliet13}, FFmpeg, Esh & S06 & Synthetic and real binary (detection) \\
\hline
Reveal\cite{Chakraborty2022Reveal}, FFmpeg+QEMU & S08 & Real-world C/C++ source (detection) \\
\hline
BigVul\cite{Fan2020BigVul} & S21 & Labeled C/C++ source, fine-grained localization \\
\hline
Linux Kernel & S15, S21 & Real-world C source code \\
\hline
Program-, firmware-, hardware-, or domain-specific (non-standard) & S01--S05, S09--S12, S14, S16--S20 & Custom/proprietary programs, RISC-V, embedded firmware, ICS protocols \\
\hline \hline
\end{tabular}%
}
\end{table}

\subsection{Research Questions}
\subsubsection{RQ1.- Application of RL in Cybersecurity Tasks}
The corpus reveals that RL is applied to a spectrum of security tasks that can be organized into a taxonomy with two groups. The first, \emph{exploration-oriented}, comprises mutation and operator selection, seed scheduling, program-path exploration, and test generation. The second, \emph{detection-oriented}, comprises both vulnerability classification and fine-grained, statement-level localization. In this context, most studies fall in the first group, confirming the main use of Reinforcement Learning.

Within the exploration-oriented group, Chen\cite{Chen2021} formalizes fuzzing as a Markov Decision Process in which the agent selects mutation operators that maximize path exploration. RiverFuzzRL\cite{Paduraru2021} casts the fuzzer as the agent that mutates program inputs to increase code coverage, and Kuznetsov et al.\cite{Kuznetsov2019,Kuznetsov20191} reduce the task to optimal mutation selection, reporting roughly 30\% faster bug discovery than random baselines.

The detection-oriented group, though smaller, is the most relevant to this review. ProRLearn\cite{article} demonstrates vulnerability detection as an RL-assisted classification problem, by combining prompt tuning with a policy gradient over C/C++ source at function level. DeepEXE \cite{Li2024} integrates RL as a decision component within a deep equilibrium network over binary CFGs, reaching state-of-the-art detection on Juliet \cite{Juliet13}, NDSS18, and FFmpeg. Finally, RLFD \cite{Jiang2025} reformulates detection as a sequential decision process and, unlike prior work, localizes vulnerabilities at the statement level, introducing a fine-grained subtask not previously present in the corpus.

This taxonomy makes the corpus structure explicit, as exploration-oriented tasks absorb the majority of studies, while the detection-oriented group is addressed by only three works, mirroring the Core/Peripheral split in Table \ref{tab:estudios-seleccionados}. This shows that RL extends beyond exploration toward source-code classification and localization, the latter remaining the least populated and still underexplored region of the task space.

\subsubsection{RQ2.- RL Algorithms and Agent Architectures}

The most widely adopted method is DQN and its variants\cite{Chen2021,Gotz2025,Huang2026,Kuznetsov2019,Kuznetsov20191,Liang2022,Paduraru2020,Paduraru2021}, in eight studies. Its dominance follows from the structure of fuzzing: mutation, seed, and operator selection define a discrete, low-dimensional action space, the regime where value-based methods are most effective. Within this group, RLF\cite{Liang2022} formalizes directed fuzzing as an MDP solved via Deep Q-Learning, DRL-Fuzzer\cite{Huang2026} adds an adaptive bit-flip variant (LABM-DQN), and RLFuzz\cite{Gotz2025} offers the most systematic comparison across DQN variants.

Policy-based and bandit methods occupy complementary niches. PPO (three studies) is chosen where the policy must be optimized under richer reward signals: Steenhoek et al.\cite{Steenhoek2025} apply it within RLHF with static-quality rewards (up to 23\% better testing), and Khan et al.\cite{Khan2025} as a mutation planner for embedded systems. UCB/MAB (about six studies) is preferred when the problem reduces to selecting among a fixed set of arms under an exploration--exploitation trade-off, without a full state representation: DocFuzz\cite{Xie2024} adopts UCB over greedy, $\epsilon$-greedy, and Thompson sampling. Wang et al.\cite{Wang2021} use UCB1 for seed scheduling, Pham et al.\cite{Pham2024} for multilevel mutation, and Syzballer\cite{Huang2022}, MARL\cite{Jhang2024}, and SAFuzz\cite{Yu2024} use MAB for task scheduling, argument selection, and operator selection.

Policy-gradient methods, in turn, lean toward detection, since these tasks optimize a policy emitting sequential or conditional label predictions rather than choosing among discrete operators. REINFORCE drives detection in DeepEXE\cite{Li2024} and RLFD\cite{Jiang2025} (the latter with a trajectory-level reward for statement-level localization), while ProRLearn\cite{article} uses Vanilla Policy Gradient with CodeBERT for classification. The result is a clear problem--algorithm mapping: value-based for discrete exploration, bandits for arm selection, PPO for stable optimization under richer rewards, and policy gradients as the natural substrate for the source-code detection and localization this review targets.

\subsubsection{RQ3.- RL Problem Formulation}

The definition of the core RL components (state, action, reward, and environment) varies markedly with the task. Table \ref{tab:formulaciones} summarizes representative formulations across the corpus and assesses, how strongly the problem structure justifies the use of RL in each case.

\begin{table}[ht]
\centering
\footnotesize
\caption{Representative RL formulations and their structural suitability}
\label{tab:formulaciones}
\resizebox{\linewidth}{!}{%
\begin{tabular}{p{2cm} p{3cm} p{3cm} p{2.8cm} p{2.2cm} p{3cm}}
\hline \hline
\textbf{Task} & \textbf{State} & \textbf{Action} & \textbf{Reward} & \textbf{Environment} & \textbf{RL suitability} \\
\hline \hline
Coverage-guided fuzzing (S01) & Execution info (buffer, block coverage, sanitizer) & Mutation/operator selection & Coverage gain, new crashes & Program under test & High (sequential, dense reward) \\
\hline
Seed/task scheduling (S15) & CFG-derived block weights, coverage & Seed/task selection & New-path or coverage gain & Kernel under fuzzing & Bandit-level (reduces to MAB) \\
\hline
Concolic exploration (S12) & Path constraints (LSTM) & Branch/path choice & New-path coverage & Binary execution & High (sequential) \\
\hline
Detection, binary (S06) & Program state per CFG transition & Branching at conditional node & DEQ convergence to correct label & Binary CFG & Moderate (sequential but label-driven) \\
\hline
Detection, source (S08) & Prompt-tuned code embedding & Vulnerable / non-vulnerable label & $+1$ if label matches, else $0$ & Labeled dataset & Low (horizon 1, $\approx$ supervised) \\
\hline
Localization, source (S21) & Line embedding + prior-prediction context & Per-line relevance (binary) & Trajectory IoU + ranking penalty & Labeled source dataset & High (sequential, delayed reward) \\
\hline \hline
\end{tabular}%
}
\end{table}

This comparison exposes that not every formulation justifies RL to the same degree. In ProRLearn \cite{article}, detection is cast as a single-step decision whose reward is merely whether the predicted label matches the ground truth. With a horizon of one and no state transitions, the formulation effectively reduces to supervised classification optimized by a policy gradient, so RL contributes little beyond the learning rule. Scheduling formulations such as Syzballer \cite{Huang2022} and Wang et al. \cite{Wang2021} reduce to selecting among a fixed set of arms and are adequately captured by multi-armed bandits, which is precisely why they adopt them instead of full RL. Genuine RL value emerges where decisions are sequential, the next state depends on the previous one, and the reward is delayed or defined over a trajectory: coverage-guided fuzzing and concolic exploration meet these conditions, and, most relevant here, RLFD \cite{Jiang2025} meets them for detection, accumulating per-line decisions under a trajectory-level reward (IoU plus a ranking penalty).

This distinction is central to the present review: a formulation in which the state is a structural position within a statically extracted CFG, and the agent performs a sequence of navigation and decision steps under a delayed localization reward, is exactly the setting where RL is structurally justified rather than decorative. This observation directly motivates the CFG-based agent identified as the research opportunity.

\subsubsection{RQ4.- Code Representations}

As detailed in the descriptive analysis and in Table \ref{tab:representaciones-codigo}, structural representations extracted directly from source code are scarce. The most structured representations in the corpus are CFGs, yet they are obtained almost exclusively from binaries, an example of this being DeepEXE\cite{Li2024}, which reconstructs them with IDA Pro, while the single source-level CFG, derived by Syzballer\cite{Huang2022} with SVF, is used to guide fuzzing. The studies that do target source-code detection rely on non-structural representations: being token sequences encoded with CodeBERT\cite{article}, and statement embeddings with CodeBERT-HLS\cite{Jiang2025}. Within the corpus, source-code detection is not approached through a statically extracted graph.

This pattern reveals a disconnect between two research lines that have evolved in parallel. On one side, source-code vulnerability detection has embraced structural and semantic representations, like ASTs and CFGs \cite{Chakraborty2022Reveal}, but predominantly in supervised non-RL settings. On the other, RL in security has operated on execution traces, coverage signals, binaries and fuzzing, rather than on static source structure. The research opportunity lies at their intersection: no study defines the agent's state as a structural representation statically extracted from source code while training a policy to navigate that structure and localize the node responsible for a vulnerability.

\subsubsection{RQ5.- Datasets, Benchmarks and Metrics}

The benchmarks in the corpus (Table \ref{tab:datasets-benchmarks}) measure different tasks, limiting comparability. LAVA-M \cite{DolanGavitt2016LAVA} and DARPA CGC \cite{DARPA_CGC} support fuzzing evaluation through known or injected binary faults but do not assess source-code detection. Juliet \cite{Juliet13} offers broad CWE coverage but may encourage overfitting to synthetic patterns, and Reveal \cite{Chakraborty2022Reveal}, FFmpeg+QEMU, and BigVul \cite{Fan2020BigVul}, though closer to the target task, still face label noise, class imbalance, and coarse granularity for fine-grained localization.

Evaluation rigor is also uneven. Few studies report cross-validation, cross-project evaluation, or generalization to unseen CWE types, so reported gains may not hold beyond the dataset measured. This is especially limiting for node-level localization, which needs fine-grained ground truth that only BigVul approaches, and with marginal adoption.

Three metric families are used. Coverage-oriented metrics (code and branch coverage, mutations to first error) prevail in fuzzing, classification metrics (Accuracy, Precision, Recall, F1) dominate detection and test-quality studies, and localization metrics appear only in RLFD \cite{Jiang2025}, namely intersection-over-union (IoU) and Top-k\%.

\subsubsection{RQ6.- Limitations, Risks and Gaps}

Synthesizing the previous questions, the corpus exhibits six recurring gaps. \emph{Methodological:} few studies present an explicit MDP formulation, and several reduce to supervised classification or to a bandit, so RL is not always structurally justified (RQ3). \emph{Representation:} the use of static structural sources as the agent state is minimal, with Syzballer \cite{Huang2022} and RLFD \cite{Jiang2025} offering only partial coverage of the target formulation. \emph{Evaluation:} benchmarks are heterogeneous and weakly comparable, mixing binary fuzzing corpora with source-level detection datasets (RQ5). \emph{Generalization:} few works report cross-project or unseen-CWE evaluation, leaving transferability uncertain, as acknowledged for fine-grained detection \cite{Jiang2025}. \emph{Baselines:} comparisons against strong supervised detectors like graph neural networks and Transformer models such as Devign \cite{Zhou2019Devign} are scarce, hindering positioning against the state of the art. \emph{Reproducibility:} the limited release of code, data splits, and training configurations hampers independent verification. Together, these gaps indicate that RL-based source-code vulnerability detection, and especially fine-grained localization, remains an emerging area with clear opportunities for future work.

\section{Discussion}
Unlike prior secondary studies, this review adopts an RL-formulation perspective. Table \ref{tab:related-reviews} contrasts the closest reviews: existing work covers static code analysis without RL or deep-learning-based fuzzing without dissecting the RL problem, whereas this review analyzes how vulnerability-analysis tasks are cast as RL problems (state, action, reward, environment) and identifies the unaddressed source-level CFG-state gap.

\begin{table}[ht]
\centering
\caption{Positioning against related secondary studies}
\label{tab:related-reviews}
\resizebox{0.8\linewidth}{!}{%
\begin{tabular}{l l c c c}
\hline \hline
\textbf{Review} & \textbf{Focus} & \textbf{RL} & \textbf{Code analysis} & \textbf{S--A--R analysis} \\
\hline \hline
Gomes et al.\cite{Gomes2025} & Static analysis (IoT security) & No & Yes & No \\
\hline
Miao et al.\cite{Miao2022} & Deep learning for fuzzing & Partial & Partial & No \\
\hline
This review & RL for C/C++ vulnerability analysis & Yes & Yes & Yes \\
\hline \hline
\end{tabular}%
}
\end{table}

The results of this SLR reveals two research streams: one optimizes exploration, mutation selection, and test generation to maximize coverage, while the other applies policy-gradient methods directly to detection over source code, as shown by the inclusion of studies by Ren et al\cite{article} and Jiang et al.\cite{Jiang2025}. A question may arise as to why RL would be competitive with the dominant supervised line based on Transformers, graph neural networks, and code encoders such as CodeBERT \cite{Chakraborty2022Reveal,Zhou2019Devign}. The evidence indicates that RL does not outperform these models at function-level classification, where strong supervised baselines are well established. Its value is therefore complementary rather than substitutive: the scientific opportunity is to pair powerful structural and semantic representations (CFGs, ASTs, code embeddings) with agents that learn navigation, prioritization, and localization policies, using RL as a decision layer over a structural encoder rather than replacing it.

This sharpens the representation gap noted earlier, as the corpus is split between binary- and source-based approaches, with almost no statically extracted source structure used as the agent state, pointing to an opportunity to unite structural static analysis with reward-driven decision-making.

\section{Conclusions}

This systematic literature review synthesized the existing evidence on the application of reinforcement learning to software vulnerability analysis and detection. The results show that its main use is concentrated in areas such as \textit{fuzzing}, test generation, and program exploration. However, recent studies demonstrate its direct application to vulnerability detection, both in binary representation and source code.

The evidence suggests that vulnerability detection through \textit{Reinforcement Learning} is technically plausible, although not yet mature, as a research line. Likewise, a limited diversity of code representations was identified, along with a scarce adoption of standardized benchmarks. These conclusions should be read in light of the review's own limitations: a relatively small corpus, a dependence on the selected databases, the heterogeneity of the tasks considered, and the small number of studies strictly aligned with static source-code analysis.

The main gap is observed in the absence of approaches that combine RL with structural representations statically obtained from source code, such as ASTs, as the state of the \textit{Reinforcement Learning} agent. RLFD constitutes the closest antecedent, demonstrating that reinforcement learning, formulated as a sequential decision-making process, improves fine-grained vulnerability localization in source code. However, it still relies on a state based on statement embeddings, leaving open the design of an RL agent that operates on a statically extracted source-code CFG to localize the node responsible for a vulnerability. Such an agent would combine the structural representations widely used in vulnerability detection with the reward-driven adaptivity of RL, a direction that remains scarcely explored.

\bibliographystyle{splncs04}
\nocite{*}
\begingroup
\raggedright
\bibliography{citas}
\endgroup
\end{document}